\documentclass[pra,twocolumn,showpacs,showkeys]{revtex4-1}
\usepackage{amsfonts}

\usepackage{graphicx}
\usepackage{dcolumn}
\usepackage{bm}
\include{biblio}

\begin{document}

\title{Intrinsic nonlinear response of surface plasmon polaritons}
\author{Song-Jin Im}
\email{ryongnam7@yahoo.com}
\author{Gum-Song Ho}
\author{Gum-Hyok Kim}
\affiliation{Department of Physics, \textbf{Kim Il Sung} University, Daesong District, Pyongyang, DPR Korea}

\begin{abstract}
We offer a model to describe the intrinsic nonlinear response of surface plasmon polaritons (SPPs). Relation of the complex nonlinear coefficient of SPPs to the third-order nonlinear susceptibility of the metal is provided. As reported in a recent study, gold is highly lossy and simultaneously highly nonlinear due to interband absorption and interband thermo-modulation at a wavelength shorter than 700 nm. The effect of the high loss of the metal on the SPP nonlinear propagation is taken into account in our model. With the model we show difference in sign of real and imaginary parts between the nonlinear propagation coefficient and the nonlinear susceptibility of component material for the first time to our knowledge. Our model could have practical importance in studying plasmonic devices utilizing the nonlinear phase modulation and the nonlinear absorption of SPPs. For example, it allows one to extract the complex nonlinear susceptibility of gold through a measurement of SPP nonlinear propagation at the visible range.
\end{abstract}

\maketitle

\section{Introduction}

Nonlinear optics and nonlinear fiber optics have led to realization of modern photonic functionalities such as optical frequency conversion, generation of ultrashort pulses, all-optical signal processing and ultrafast switching \cite%
{Boyd_academic_2008,Agrawal_academic_2007}. In the last decades, many efforts have been to the study of nonlinear optical processes assisted by plasmonic nanostructures \cite%
{Kauranen _naturep_2012} because of nanoscale field confinement, field enhancement at metal-dielectric interfaces \cite%
{Barnes_nature _2003,Schuller_naturem_2010}, a strong and ultrafast third-order nonlinear response of metal \cite%
{Hache_app_1988} and other novel phenomena not achievable in conventional structures, for example extreme-ultraviolet continuum generations \cite%
{Husakou_springer_2015}.

 For extended metal-dielectric interfaces, surface electromagnetic waves can propagate at the metal-dielectric interfaces, which are called surface plasmon polartons (SPPs). SPP-mediated nonlinear processes have been extensively studied, for example harmonic generation \cite%
{Grosse_physl_2012}, 
four-wave mixing \cite%
{Renger_physl_2010}, 
nonlinear absorption \cite%
{Rotenberg_physb_2007}, 
self-phase modulation \cite%
{Davoyan_optxp_2008,Degiron_physa_2010,Skryabin_optsc_2011,Marini_physa_2011} 
and soliton \cite%
{Feigenbaum_optl_2007,Liu_physl_2007,Ye_physl_2010,Marini_optxp_2011,Marini_prl_2013}
 and all-optical modulation \cite%
{MacDonald_naturep_2008}. 

Many theoretical studies of the nonlinear response of SPPs in different structures of plasmonic waveguides have been performed \cite%
{Davoyan_optxp_2008,Degiron_physa_2010,Conforti_physb_2012,Marini_njp_2013,
Leon_physa_2014,Skryabin_optsc_2011,Marini_physa_2011,
Feigenbaum_optl_2007,Liu_physl_2007,Ye_physl_2010,Marini_optxp_2011,Ginzburg_optl_2010,Ginzburg_njp_2013,Marini_prl_2013}. A common theoretical approach to describe the nonlinear response in optical waveguides and fibers is based on the scalar model \cite%
{Agrawal_academic_2007}
 disregarding influence of a longitudinal component of electric field which is justified for large core waveguides. Full vectorial model \cite%
{Afshar_optxp_2009,Afshar_optxp_2013}
 is developed for describing the nonlinear response in subwavelength waveguides. In \cite%
{Skryabin_optsc_2011,Marini_physa_2011}
 the authors offered approaches to describe a surface-induced enhancement of the nonlinear response in plasmonic planar waveguides \cite%
{Skryabin_optsc_2011}
and rod waveguides \cite%
{Marini_physa_2011}, 
which are in good agreement with the full vectorial model \cite%
{Afshar_optxp_2009}. In most of these studies \cite%
{Davoyan_optxp_2008,Degiron_physa_2010,Skryabin_optsc_2011,Marini_physa_2011,
Feigenbaum_optl_2007,Liu_physl_2007,Ye_physl_2010,Marini_optxp_2011,Marini_prl_2013}, a linear metal and a nonlinear dielectric were assumed and  an imaginary part of permittivity of the metal $\varepsilon''_{\tt{m}}$ is assumed to be small and perturbative.  However, in practice the metal shows a strong nonlinear response \cite%
{Hache_app_1988}, and SPPs are intrinsically nonlinear even in the absence of a nonlinear dielectric. 

The nonlinear response of SPP based on the ponderomotive nonlinearity of metals in the infrared spectral range \cite%
{Ginzburg_optl_2010} and the second-order surface plasmon solitons based on the non-local multipole nonlinearities in metal surface \cite%
{ Ginzburg_njp_2013} were shown.

Marini et al. \cite%
{Marini_njp_2013} studied the effects of the thermo-modulational interband nonlinearity of gold on the propagation of SPPs guided on gold nanowires at the wavelengths longer than 750 nm. They also assumed low loss of gold which is justified at the long wavelengths. However, at shorter wavelengths gold has a high loss \cite%
{Johnson_physb_1972} and thus an effect of the high loss of gold is needed to be considered for modeling SPP nonlinear propagation. Moreover, the quite large third-order nonlinear susceptibility of gold at these shorter wavelengths, as recently reported in \cite%
{Conforti_physb_2012,Marini_njp_2013}, 
gives practical importance of modeling the SPP nonlinear propagation in the spectral range \cite%
{Leon_physa_2014,Leon_optl_2014}. De Leon et al. ~\cite%
{Leon_physa_2014}
developed an approach to describe the intrinsic nonlinear response of SPPs in planar waveguides taking account of the significant loss of gold $\varepsilon''_{\tt{m}}$. They assumed only that the SPP loss coefficient is not too large. It is worthy to note that the condition of small SPP loss coefficient is not equivalent to the condition of small $\varepsilon''_{\tt{m}}$  and even in the case of large $\varepsilon''_{\tt{m}}$  the SPP loss coefficient can be small if the power portion in metal is small or the the SPP loss is compensated with a gain media in dielectrics \cite%
{Berini_naturep_2011}. However, this approach can be applied in planar waveguides, thus a  general approach is needed.

In this paper we offer a model to describe the intrinsic nonlinear response of SPPs in a spectral range where the plasmonic metal is highly lossy, which is a modified version of the full-vectorial model. Our model shows a prominent difference in phase angle (one can find a hint of the difference in phase angle in Fig.~\ref{fig:3} of \cite%
{Leon_physa_2014})
and even in sign between the complex third-order nonlinear susceptibility of the metal and the complex effective nonlinear coefficient of SPPs. In particular, a positive imaginary part of the complex nonlinear susceptibility of the metal, corresponding to a positive nonlinear absorption, can result in a negative nonlinear absorption (saturated absorption) of SPPs. We note that our approach can be used for all types of waveguides including the planar waveguides of \cite%
{Leon_physa_2014}.

\section{Theoretical model}

In this paper we assume the continuous wave and all the fields can be expressed as  $\vec{\tt{F}}\left(\vec{r},t\right)=\left(1/2\right)\cdot\left[\vec{F}\left(\vec{r}\right)
{\tt{exp}}
\left(-i\omega t\right)+c.c\right]$, where c.c. signifies the complex conjugate. Below, we consider the time-independent fields 
$\vec{F}\left(\vec{r}\right)$. For a linear plasmonic waveguide, we can write the electric and magnetic field vectors
\begin{eqnarray}
	\vec{e}\left(\vec{r}\right)=\sqrt{Z_{0}/s_{0}}\Psi_{0}\left(z\right){\tt{exp}}
\left(ikz\right)\vec{e_{0}}\left(\vec{r_{\bot}}\right),
\label{eq:1}	
\end{eqnarray}
 \begin{eqnarray}
	\vec{h}\left(\vec{r}\right)=\left(i\omega \mu_{0}\right)^{-1}\nabla\times\vec{e}=\nonumber\\
	\sqrt{1/(Z_{0}s_{0})}\Psi_{0}\left(z\right)
{\tt{exp}}\left(ikz\right)\vec{h_{0}}\left(\vec{r_{\bot}}\right).
\label{eq:2}
\end{eqnarray}
Here $\varepsilon_{0}$, $\mu_{0}$   
and 
 $Z_{0}=\sqrt{\mu_{0}/\varepsilon_{0}}$
 is the permittivity, permeability and wave impedance in vacuum, respectively.  $z$ and $\hat{z}$  are a coordinate and the unit vector in the direction of propagation and  $\vec{r_{\bot}}$ 
is a position vector in the transverse plane.  $s_{0}$ is defined as 
$s_{0}=\left(1/2\right)\int{{\tt{Re}}\left(\vec{e_{0}}
\times\vec{h^{*}_{0}}\right)\cdot\hat{z}d\sigma}$, 
where the integral is performed in the transverse plane. The function $\Psi_{0}\left(z\right)$  has the form  
$\Psi_{0}\left(z\right)=\Psi_{0}\left(0\right){\tt{exp}}\left(-\alpha z/2\right)$
and $\beta=\kappa+i\alpha /2$  is the SPP propagation constant. We note that the  $\Psi_{0}$ is normalized so that the  $\left|\Psi_{0}\right|^{2}$ is equal to the power flow along the $z$ direction 
$S\left(z\right)=\left(1/2\right)\int{{\tt{Re}}\left(\vec{e}
\times\vec{h^{*}}\right)\cdot\hat{z}d\sigma}=\left|\Psi_{0}\right|^{2}$.

Now, we add the small perturbation of permittivity $\delta\varepsilon$  to the linear plasmonic waveguide. If we assume a transverse distribution of the perturbation 
$\delta\varepsilon=\delta\varepsilon\left(\vec{r_{\bot}}\right)$  and a longitudinal invariance of the perturbation 
$d\left(\delta\varepsilon\right)/dz=0$, the perturbed electric and magnetic fields can be expressed as follows.
\begin{eqnarray}
	\vec{E}\left(\vec{r}\right)=\sqrt{Z_{0}/s_{0}}\Psi\left(z\right){\tt{exp}}
\left(ikz\right)\nonumber\\
\times\left[\left(\vec{e_{0}}\left(\vec{r_{\bot}}\right)+\delta\vec{e_{0}}
\left(\vec{r_{\bot}}\right)\right)\right],
\label{eq:3}	
\end{eqnarray}
 \begin{eqnarray}
	\vec{H}\left(\vec{r}\right)=\left(i\omega \mu_{0}\right)^{-1}\nabla\times\vec{E}=\nonumber\\
	\sqrt{1/Z_{0}s_{0}}
\Psi\left(z\right){\tt{exp}}\left(ikz\right)\left[\left(\vec{h_{0}}\left(\vec{r_{\bot}}\right)+
\delta\vec{h_{0}}\left(\vec{r_{\bot}}\right)\right)\right].
\label{eq:4}
\end{eqnarray}

Let's remind of the Lorentz reciprocity theorem \cite%
{Snyder_hall_1983,Ye_physl_2010}.
 \begin{eqnarray}
	\frac{\partial}{\partial z}\int \left[\vec{E_{1}}\left(\vec{r}\right)\times\vec{H_{2}}\left(\vec{r}\right)-
\vec{E_{2}}\left(\vec{r}\right)\times\vec{H_{1}}\left(\vec{r}\right)\right]
\cdot\hat{z}d\sigma=\nonumber\\
i\frac{k_{0}}{Z_{0}}\int\left[\varepsilon_{2}\left(\vec{r}\right)-\varepsilon_{1}
\left(\vec{r}\right)\right]\vec{E_{1}}\left(\vec{r}\right)\cdot\vec{E_{2}}
\left(\vec{r}\right)d\sigma,
\label{eq:5}
\end{eqnarray}
where 
$\left(\vec{E_{1}},\vec{H_{1}}\right)$
  and 
$\left(\vec{E_{2}},\vec{H_{2}}\right)$
  are solutions of the Maxwell equations corresponding to the relative permittivity $\varepsilon_{1}\left(\vec{r}\right)$  
and 
$\varepsilon_{2}\left(\vec{r}\right)$, 
respectively. Let's choose 
$\left(\vec{E_{1}},\vec{H_{1}}\right)$
  and 
$\left(\vec{E_{2}},\vec{H_{2}}\right)$
  to be the unperturbed backward propagating field 
$\left(\vec{e}^{-},\vec{h}^{-}\right)$
  and the perturbed forward propagating field 
$\left(\vec{E},\vec{H}\right)$. 
In the first order of perturbation, Eq.~(\ref{eq:5}) leads to the following equation describing the amplitude $\Psi\left(z\right)$.
\begin{eqnarray}
\frac{d\Psi}{dz}=-\frac{\alpha}{2}\Psi+ik_{0}\cdot\delta n_{\tt{eff}}\Psi,
\label{eq:6}	
\end{eqnarray}
 \begin{eqnarray}
	\delta n_{\tt{eff}}=\frac{\int\delta\varepsilon\cdot\left(\vec{e_{0}}^{2}-2e_{0z}^{2}\right)d\sigma}{2Z_{0}\int\left(\vec{e_{0}}\times\vec{h_{0}}\right)\cdot\hat{z}d\sigma}
\label{eq:7}
\end{eqnarray}
Here we considered  $e_{0x}^{-}=e_{0x}$, $e_{0y}^{-}=e_{0y}$,  $e_{0z}^{-}=-e_{0z}$,   $h_{0x}^{-}=-h_{0x}$, $h_{0y}^{-}=-h_{0y}$,  $h_{0z}^{-}=h_{0z}$.

We emphasize that in the derivation of Eq.~(\ref{eq:6}) and (\ref{eq:7}) from Eq.~(\ref{eq:3})--(\ref{eq:5}), no additional assumption has been taken, except for the small perturbation and longitudinal invariance of $\delta\varepsilon\left(\vec{r_{\bot}}\right)$. In particular, the assumption of small imaginary part of metal permittivity have never been taken, contrary to in the full-vectorial \cite%
{Skryabin_optsc_2011,Marini_physa_2011,Afshar_optxp_2009,Afshar_optxp_2013}
 and scalar \cite%
{Agrawal_academic_2007}
 models.   In our model the backward propagating fields are used as the multiplied ones and consequently the product terms on the left-hand side of  Eq.~(\ref{eq:5}) experience no loss and the imaginary part of metal permittivity on the right-hand side also disappears, while in the full-vectorial and scalar models conjugated fields were multiplied. We also note that in \cite%
{Ye_physl_2010} a formalism utilizing backward propagating fields was provided.

\begin{figure*}
\includegraphics[width=0.7\textwidth]{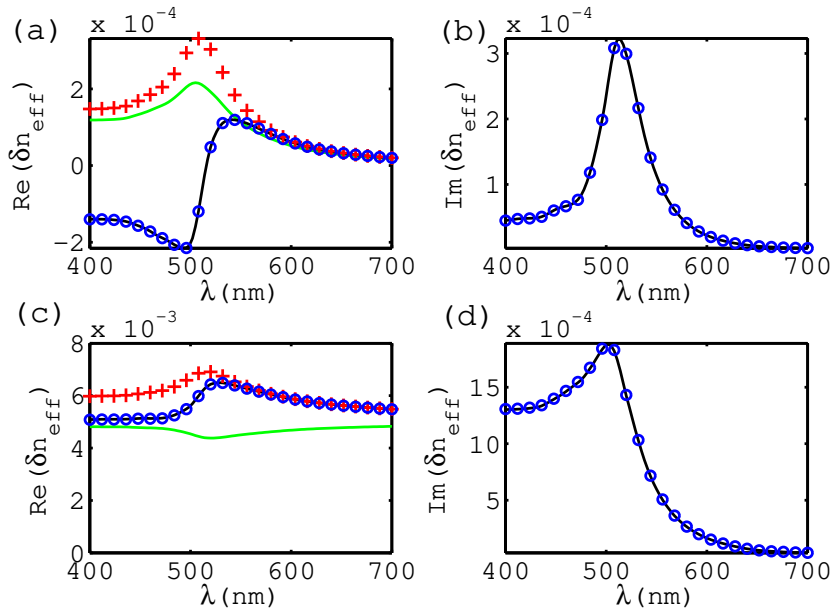}
\caption{(Color online) Response of effective refractive index in the single gold-air interface to the perturbation of permittivity   in the metal (a), (b) and the dielectric (c), (d), respectively. Real part (a), (c) and imaginary part (b), (d) of the response. The black line corresponds to our model (\ref{eq:7}), the blue circles to the analytical formula (\ref{eq:8}), the red crosses to the full-vector model (\ref{eq:9}) and the green line to the scalar model (\ref{eq:10}). }
\label{fig:1}
\end{figure*} 

To justify the model (7), we utilize this model in the single gold-air interface  ($\varepsilon=\varepsilon_{\tt{gold}}\cdot\theta\left(-x\right)$ + $\delta\varepsilon_{\tt{air}}\cdot\theta\left(x\right)$) 
for 
$ \delta\varepsilon=\delta\varepsilon_{1}\cdot\theta\left(-x\right)+\delta\varepsilon_{2}\cdot\theta\left(x\right)$
, where 
$\theta\left(x\right)$  is the Heaviside step function, and compare with an analytical result, the full-vectorial model and the scalar model. For calculations we used the measured wavelength-dependant permittivity of gold \cite%
{Johnson_physb_1972}. If we use the formula of effective refractive index 
$n_{\tt{eff}}=\sqrt{\varepsilon_{1}\varepsilon_{2}/\left(
\varepsilon_{1}+\varepsilon_{2}\right)}$
  in the single interface \cite%
{Maier_springer_2007}, we can obtain the analytical formula of  $\delta n_{\tt{eff}}$  as following
 \begin{eqnarray}
	\delta n_{\tt{eff}}=\sqrt{\frac{\left(
\varepsilon_{1}+\delta\varepsilon_{1}\right)\left(
\varepsilon_{2}+\delta\varepsilon_{2}\right)}{\left(
\varepsilon_{1}+\delta\varepsilon_{1}\right)+\left(
\varepsilon_{2}+\delta\varepsilon_{2}\right)}}-\sqrt{\frac{\varepsilon_{1}\varepsilon_{2}}{\varepsilon_{1}+\varepsilon_{2}}}.
\label{eq:8}
\end{eqnarray}

By the full-vectorial model \cite%
{Skryabin_optsc_2011,Afshar_optxp_2009}
 and the scalar model \cite%
{Agrawal_academic_2007}, $\delta n_{\tt{eff}}$ can be expressed
\begin{eqnarray}
	\delta n_{\tt{eff}}=\frac{\int\delta\varepsilon\cdot\left|\vec{e_{0}}\right|^{2}d\sigma}{2Z_{0}\int{\tt{Re}}\left(\vec{e_{0}}\times\vec{h_{0}^{*}}\right)\cdot\hat{z}d\sigma},
\label{eq:9}
\end{eqnarray}
 \begin{eqnarray}
	\delta n_{\tt{eff}}=\frac{k_{0}}{2\kappa}\frac{\int\delta\varepsilon\cdot\left|\vec{e_{0}}\right|^{2}d\sigma}{\int\left|\vec{e_{0}}\right|^{2}d\sigma}.
\label{eq:10}
\end{eqnarray}

First we calculate for the case of 
$\delta\varepsilon_{1}=0.01$, $\delta\varepsilon_{2}=0$
  [Fig.~\ref{fig:1}(a) and (b)], where the perturbation is taken in the metal. As shown in Fig.~\ref{fig:1}, the result calculated by our model (\ref{eq:7}) coincides with that by the analytical formula (\ref{eq:8}) in all the considered range of wavelength. However, it greatly differs from results of the full-vectorial (\ref{eq:9}) and scalar (\ref{eq:10}) models at wavelengths shorter than 600 nm showing even difference in sign [see Fig.~\ref{fig:1}(a)] because the large imaginary part of permittivity of gold at these wavelengths influences a phase of the electric field in metal which results in difference in value of denominator between Eq.~(\ref{eq:7}) and Eq.~(\ref{eq:9}) and (\ref{eq:10}). And our model shows an imaginary part of $\delta n_{\tt{eff}}$  produced by the real perturbation $\delta\varepsilon$  which can not be described by the full-vector and scalar models [see Fig.~\ref{fig:1}(b)].

For the case of 
$\delta\varepsilon_{1}=0$, $\delta\varepsilon_{2}=0.01$, 
where the perturbation is taken in the dielectric, the result of our model is almost same as that of the full-vector model [see Fig.~\ref{fig:1}(c)] because even the large imaginary part of permittivity of gold doesn't greatly influence the electric field in dielectric.

\section{Intrinsic nonlinear response of surface plasmon polaritons}

To investigate an intrinsic nonlinear response of SPPs, we take the nonlinear perturbation $\delta\varepsilon=\varepsilon_{\tt{\tt{nl}}}$. If we take the self-nonlinear permittivity 
$\varepsilon_{\tt{nl,self}}=\left(3/4\right)\chi^{\left(3\right)}\left(\omega;\omega,-\omega,\omega\right)\left|\vec{E}\right|^{2}$
 \cite%
{Boyd_academic_2008,Agrawal_academic_2007,Leon_physa_2014}
, Eq.~(\ref{eq:6}) and (\ref{eq:7}) lead to 
\begin{eqnarray}
\frac{d\Psi}{dz}=-\frac{\alpha}{2}\Psi+i\gamma\left|\Psi\right|^{2}\Psi,
\label{eq:11}	
\end{eqnarray}
\begin{widetext}
 \begin{eqnarray}
	\gamma=k_{0}\frac{\left(3/4\right)\int\chi^{\left(3\right)}\left|\vec{e_{0}}\right|^{2}\cdot\left(\vec{e_{0}}^{2}-2e_{0z}^{2}\right)d\sigma}{Z_{0}\int\left(\vec{e_{0}}\times\vec{h_{0}}\right)\cdot\hat{z}d\sigma\cdot\int {\tt{Re}}\left(\vec{e_{0}}\times\vec{h_{0}^{*}}\right)\cdot\hat{z}d\sigma},
\label{eq:12}
\end{eqnarray}
\end{widetext}
where $\gamma$  is the effective nonlinear coefficient and $\chi^{\left(3\right)}$  is the third-order nonlinear susceptibility.

\begin{figure}
\includegraphics[width=0.5\textwidth]{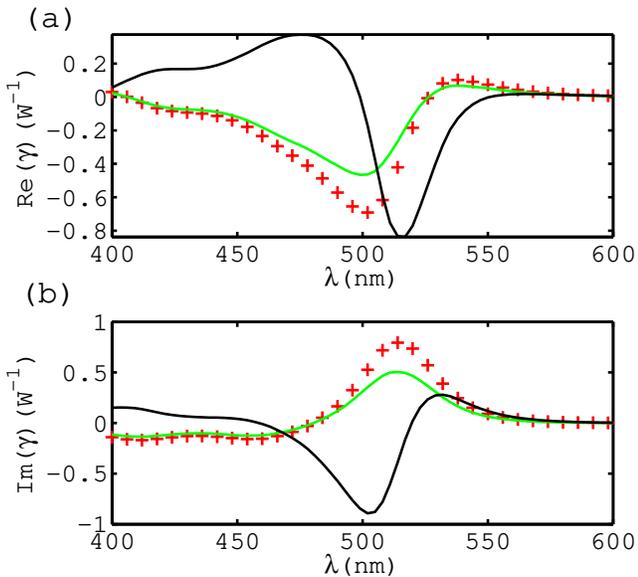}
\caption{(Color online) Effective nonlinear coefficient in the single gold-air interface according to the wavelength. The black line is by our model (\ref{eq:7}),(\ref{eq:12}), the red crosses by the full-vector model (\ref{eq:9}) and the green line by the scalar model (\ref{eq:10}). (a) and (b) shows real and imaginary part of the effective nonlinear coefficient, respectively.}
\label{fig:2}
\end{figure}

\begin{figure}
\includegraphics[width=0.5\textwidth]{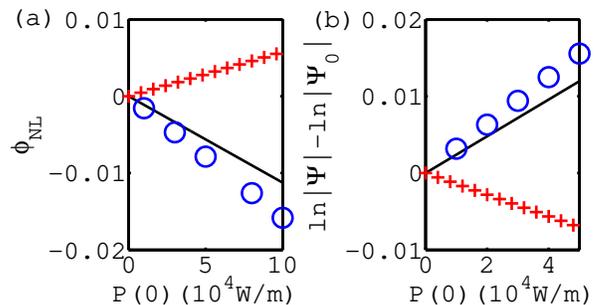}
\caption{(Color online) Power-dependence of phase shift for the wavelength of 535 nm and the propagation distance of 1000 nm (a) and absorption for the wavelength of 502 nm the propagation distance of 500 nm (b) in the single gold-air interface.  The black line is by our model, the red crosses by the full-vector model, the blue circles by a full-dimensional simulation.}
\label{fig:3}
\end{figure}

We apply this model [Eq.~(\ref{eq:11}) and (\ref{eq:12})] to the single gold-air interface. It is assumed that the effective nonlinearity results from the intrinsic self-nonlinearity of gold, while the dielectric is taken as a linear medium. The reported self-nonlinear susceptibility of gold $\chi^{\left(3\right)}$ \cite%
{Marini_njp_2013}
 is used. As expected, Fig.~\ref{fig:2} shows that the effective nonlinear coefficient by our model significantly deviates from those by other models at wavelengths shorter than 600 nm. We can know from  Eq.~(\ref{eq:9}) and (\ref{eq:10}) that a phase angle of  the complex effective nonlinear coefficient $\gamma$ by the full-vector and scalar model  is same as one of the complex nonlinear susceptibility of gold $\chi^{\left(3\right)}$. Therefore, it is clear from Fig.~\ref{fig:2} that our model shows difference in phase angle between $\gamma$ and $\chi^{\left(3\right)}$, even sign-difference in real and imaginary parts of them. In particular, we can see a negative real part of  $\gamma$ meaning a negative nonlinear phase-shift [see Fig.~\ref{fig:3}(a)] in spite of a positive real part of $\chi^{\left(3\right)}$  at the wavelength of 535 nm and a negative imaginary part of $\gamma$  meaning a negative nonlinear absorption [see Fig.~\ref{fig:3}(b)] in spite of a positive imaginary part of $\chi^{\left(3\right)}$  at the wavelength of 502 nm. Our results are in moderate agreement with results of full-dimensional simulation (blue circles of Fig.~\ref{fig:3}).
 
\begin{figure*}
\includegraphics[width=0.7\textwidth]{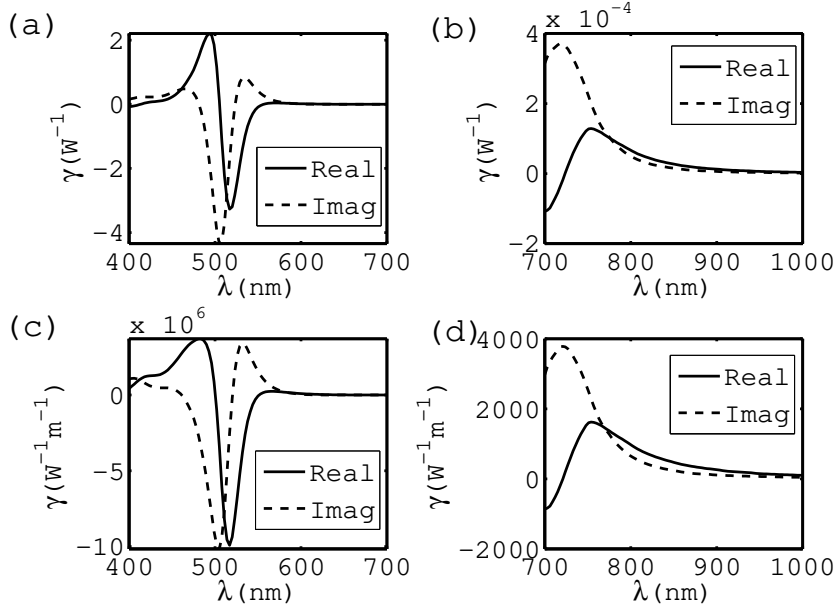}
\caption{ Effective nonlinear coefficient of antisymmetric mode in 60-nm-thick gold slab surrounded by air (a),(b) and TM mode ($m=0$) in gold rod waveguide with air cladding and the waveguide radius of 100 nm (c),(d). The solid line is real part of the effective nonlinear coefficient and the dashed line is imaginary part of that.}
\label{fig:4}
\end{figure*}

As other examples of plasmonic waveguides, 64-nm-thick gold slab surrounded by air and gold rod waveguide with air cladding and the waveguide radius of 100 nm are calculated by our model.  For calculations, the analytical results for modes in metal slab \cite%
{Maier_springer_2007} and metal rod \cite%
{Marini_physa_2011}
 are substituted to Eq.~(\ref{eq:12}). Fig.~\ref{fig:4}(a) and (b) shows the effective nonlinear coefficient in the gold slab (antisymmetric mode) which is about ten times greater than that in the single gold-air interface. Fig.~\ref{fig:4}(c) and (d) shows the effective nonlinear coefficient in the gold-rod (TM mode, $m=0$) which is about billion times greater than in conventional fibers \cite%
{Agrawal_academic_2007}  and greater by several orders of magnitude than in nonlinear nanorod waveguides utilizing nonlinear dielectrics \cite%
{Marini_physa_2011}. The enhancement of nonlinearity in the examples of plasmonic waveguides is due to the field enhancement at metal-dielectric interfaces \cite%
{Marini_physa_2011} and the strong third-order nonlinear response of metal, in particular the large interband nonlinear susceptibility of gold \cite%
{Marini_njp_2013}.

It is expected that our model can be applied to extraction of a nonlinear susceptibility of metal from experimental data. For example, in Ref.~\cite%
{Leon_physa_2014}
, the authors developed a theoretical approach to describe the effective nonlinear coefficient in planar waveguides and extracted the third-order nonlinear susceptibility of gold $\chi^{\left(3\right)}$  from a measured effective nonlinear coefficient in a gold slab at the wavelength of 796 nm which characterizes the nonlinear phase shift and nonlinear absorption experienced by the SPP \cite%
{Leon_optl_2014}. For comparison, we obtained $\gamma=\left(1.16+i1.00\right)\times10^{-7}V^{2}/m^{2}$
  by our model in the single gold-air interface at $\lambda=796$ nm  from the extracted value of 
$\chi^{\left(3\right)}=\left(4.67+i3.03\right)\times10^{-19}V^{2}/m^{2}$ \cite%
{Leon_physa_2014}
, which is in a good agreement with 
$\gamma=\left(1.03+i0.98\right)\times10^{-7}V^{2}/m^{2}$
  \cite%
{Leon_physa_2014}.

It is worthy to mention again that the accuracy of our model is limited by both the approximations used in our theory and in all the other models, the small perturbation and longitudinal invariance of $\delta\varepsilon$. For validity of the first approximation, the power should be limited so that the nonlinear permittivity is small compared to the linear permittivity. For a greater power, a nonperturbative nonlinear model \cite%
{Im_laserp_2014} should be employed. For validity of the second approximation, the condition that the effective linear absorption $\alpha$  is small compared to the effective propagation constant $\beta$  should be satisfied, which has been also used in other models including \cite%
{Marini_njp_2013,Leon_physa_2014,Skryabin_optsc_2011,Marini_physa_2011}. This condition is satisfied if the SPP mode is appropriately differ from an exact resonance as in the cases above (Fig.~\ref{fig:2}--\ref{fig:4}).

It is predicted that if one uses a cross-nonlinearity with a pumping SPP at a longer wavelength instead of the self-nonlinearity, the total nonlinear phase shift and nonlinear absorption can benefit from a longer propagation length of the pumping SPP at a longer wavelength and a large interband nonlinear susceptibility of metal \cite%
{Marini_njp_2013}
 experienced by signal SPP at a shorter wavelength. Based on the theoretical consideration \cite%
{Marini_njp_2013}
 for the self-nonlinear susceptibility, the cross-nonlinear susceptibility of metal can be expressed as following.
\begin{eqnarray}
\chi^{\left(3\right)}\left(\omega_{\tt{\tt{s}}};\omega_{\tt{\tt{p}}},-\omega_{\tt{\tt{p}}},\omega_{\tt{\tt{s}}}\right)=\frac{1}{2}\varepsilon_{0}\omega_{\tt{\tt{p}}}\varepsilon''_{\tt{\tt{m}}}\left(\omega_{\tt{\tt{p}}}\right)\gamma_{\tt{T}}\left(\omega_{\tt{\tt{s}}}\right).
\label{eq:13}	
\end{eqnarray}

If the pumping power greater than the signal power is introduced, the cross-nonlinearity is dominant in the total nonlinearity and Eq.~(\ref{eq:6}) and (\ref{eq:7}) lead to 
\begin{eqnarray}
\frac{d\Psi_{\tt{\tt{s}}}}{dz}=-\frac{\alpha_{\tt{\tt{s}}}}{2}\Psi_{\tt{\tt{s}}}+i2\gamma_{\tt{\tt{cross}}}\left|\Psi_{\tt{\tt{p}}}\right|^{2}\Psi_{\tt{\tt{s}}},
\label{eq:14}	
\end{eqnarray}

\begin{widetext}
 \begin{eqnarray}
	\gamma_{\tt{\tt{cross}}}=k_{0}\frac{\left(3/4\right)\int\chi^{\left(3\right)}\left(\omega_{\tt{s}};\omega_{\tt{p}},-\omega_{\tt{p}},\omega_{\tt{s}}\right)\left|\vec{e_{0}}\left(\omega_{\tt{p}}\right)\right|^{2}\left[\vec{e_{0}}^{2}\left(\omega_{\tt{s}}\right)-2e_{0z}^{2}\left(\omega_{\tt{s}}\right)\right]d\sigma}{Z_{0}\int\left[\vec{e_{0}}\left(\omega_{\tt{s}}\right)\times
\vec{h_{0}}\left(\omega_{\tt{s}}\right)\right]\cdot\hat{z}d\sigma\cdot\int {\tt{Re}}\left[\vec{e_{0}}\left(\omega_{\tt{p}}\right)\times\vec{h_{0}^{*}}\left(\omega_{\tt{s}}\right)\right]\cdot\hat{z}d\sigma},
\label{eq:15}
\end{eqnarray}
\end{widetext}
where  $\vec{e_{0}}\left(\omega_{\tt{k}}\right)$ and $\vec{h_{0}}\left(\omega_{\tt{k}}\right)$  is the transverse distributions of electric and magnetic fields [see Eq.~(\ref{eq:1}) and (\ref{eq:2})], respectively, at the frequency of $\omega_{\tt{k}}$.  The subscripts $s$  and $p$ denotes  the signal and pumping SPPs, respectively. The maximum nonlinear phase shift and absorption is expressed as follows.
\begin{eqnarray}
{\tt{Max}}\left(\phi_{\tt{nl}}\right)=2{\tt{Re}}\left(\gamma_{\tt{cross}}\right)\left|\Psi_{\tt{p}}\left(0\right)\right|^{2}/\alpha_{\tt{p}},
\label{eq:16}	
\end{eqnarray}
 \begin{eqnarray}
{\tt{Max}}\left({\tt{ln}}\frac{\left|\Psi_{\tt{s}}\right|^{2}}{\left|\Psi_{\tt{s0}}\right|^{2}}\right)=-4{\tt{Im}}\left(\gamma_{\tt{cross}}\right)\left|\Psi_{\tt{p}}\left(0\right)\right|^{2}/\alpha_{\tt{p}}.
\label{eq:17}
\end{eqnarray}

Finally, we would like to note that our model for the cross-nonlinear coefficient [Eq.~(\ref{eq:14}) and (\ref{eq:15})] is fully accurate even at a SPP resonance for the signal, because $\alpha_{\tt{p}}$  is small and the longitudinal invariance of $\varepsilon_{\tt{nl}}$  is fully satisfied in spite of a large $\alpha_{\tt{s}}$.

\section{Conclusions}

We provided the relation of the complex SPP nonlinear coefficient to the third-order nonlinear susceptibility of the metal. The effect of the high loss of the metal on the nonlinear propagation of SPPs was fully taken into account. Backward propagating field instead of adjoint forward propagating one was utilized to derive the relation. We predicted that the high loss of gold due to interband transition can lead to difference in phase angle between the third-order nonlinear susceptibility of gold and the SPP nonlinear coefficient and even difference in sign of real and imaginary parts between them. In particular, a negative imaginary part of the third-order nonlinear susceptibility of gold, which normally corresponds to the negative nonlinear absorption or the saturated absorption, can lead to a positive imaginary part of the SPP nonlinear coefficient, which means the positive nonlinear absorption or the reverse-saturated absorption. To our knowledge for the first time, the sign-difference between the nonlinear propagation coefficient and the nonlinear susceptibility of component material was reported. Although the high interband absorption of the metal can lead to a short SPP propagation length which wouldn't be promising for nonlinear devices, the metal has the great thermo-modulational interband nonlinearity at the interband transition wavelengths \cite%
{Marini_njp_2013}. Moreover, recently optical amplification of SPP by using gain media in component dielectrics has been demonstrated \cite%
{Berini_naturep_2011} and increasing the SPP propagation length could be a possible strategy to take advantage of the great interband nonlinearity of the metal.

\end{document}